\documentclass[%
 reprint,
superscriptaddress,
 amsmath,amssymb,
 aps,
]{revtex4-1}

\usepackage{graphicx}
\usepackage{dcolumn}
\usepackage{bm}
\usepackage{siunitx}
\usepackage{subcaption}
\usepackage{hyperref}

\begin{document}

\preprint{APS/123-QED}

\title{Mass-dependent cuts in longitudinal phase space}

\author{P. Pauli} \email{p.pauli.1@research.gla.ac.uk}
\author{D.I. Glazier}
\affiliation{SUPA School of Physics and Astronomy, University of Glasgow, Glasgow G12 8QQ, Scotland, UK}

\author{M. Battaglieri}
\author{A. Celentano}
\author{R. De Vita}
\affiliation{Instituto Nazionale di Fisica Nucleare, Sezione di Genova, 16146 Genova, Italy}

\author{S. Diehl}
\affiliation{University of Connecticut, Storrs, Connecticut 06269, USA}
\affiliation{Physics Institute, Justus-Liebig-Universit\"{a}t Giessen, 35392 Giessen, Germany}

\author{A. Filippi}
\affiliation{Instituto Nazionale di Fisica Nucleare, Sezione di Torino, 10125 Torino, Italy}

\author{J.T. Londergan}
\affiliation{Dept. of Physics, Indiana University, Bloomington, IN 47405, USA}

\author{V. Mathieu}
\affiliation{Theory Center, Thomas Jefferson National Accelerator Facility, 12000 Jefferson Avenue, Newport News, VA 23606, USA}

\author{A.~P.~Szczepaniak}
\affiliation{Dept. of Physics, Indiana University, Bloomington, IN 47405, USA}
\affiliation{Theory Center, Thomas Jefferson National Accelerator Facility, 12000 Jefferson Avenue, Newport News, VA 23606, USA}
\affiliation{Center for Exploration of Energy and Matter, Indiana University, Bloomington, IN 47403, USA}

\date{\today}

\begin{abstract}
Longitudinal phase space analyses as introduced by van Hove provided a simplified method of separating different reaction production mechanisms. 
Cuts in the longitudinal phase space can help to select specific reaction kinematics but also induce nonflat acceptance effects in angular distributions. We show that in photoproduction reactions dominated by {\it{t}}-channel exchanges, selection of meson or baryon production over a large mass range can be optimized through calculating mass-dependent cut limits compared to cuts on a van Hove plot sector alone. A cut is presented that improves this selection of one type of hadron production by rejecting another. In addition we demonstrate that using cuts in longitudinal phase space preserves sufficient information to reliably extract observables from the angular distribution of the final state particles. 
\end{abstract}

\maketitle


\section{\label{sec:introduction}Introduction}
Analyzing data coming from modern day hadron and particle physics experiments involves many steps. A critical part is the identification of the reaction of interest. Assuming that one manages to identify all particles in a final state correctly, one is often still left with the situation that different underlying processes can result in the same final state particles. An example for this is the reaction $\gamma p \rightarrow p K^+ K^-$. The final state can arise from a decaying meson (e.g. $\phi\rightarrow K^+K^-$) or baryon (e.g. $\Lambda(1520)\rightarrow pK^-$). Usually one is interested in one of the reactions at a time (signal). This leaves the other as a background that interferes with the signal and generally cannot be completely removed by applying classical cuts, e.g., a cut around the invariant mass of the signal resonance.\\ \indent
Recently the JPAC collaboration investigated $K^{+}K^{-}$ photoproduction in the  double-Regge exchange limit using a dual model based on the extension of the Veneziano amplitude \cite{shi_double-regge_2015}. By analyzing the Dalitz and van Hove plots, also known as longitudinal plots, they were able to determine phase space cuts that enhanced this contribution. A similar approach would be relevant for the ongoing Jefferson Lab experiments GlueX and CLAS12 which will be able to measure this reaction with high intensity at $7$ to \SI{10}{GeV} photon energy.
In Jefferson Lab kinematics one expects production of either meson or baryon resonances via single Regge exchange to dominate. Here the goal is to isolate either the baryon or meson resonances to determine spin observables and thus the properties of the contributing states. In this spirit van Hove cuts were previously investigated with CLAS photoproduction data, where some kinematic limitations, that are further  addressed here, were found \cite{Glazier:MesonEx}.\\ \indent
Specifically in this paper we introduce a method that helps to enhance one type of reaction over the other by using the reaction kinematics. For that we explain longitudinal phase space plots in Sec. \ref{sec:lps}. In Sec. \ref{sec:cuts} we discuss the possibility of cuts in longitudinal phase space and make comparisons between different types of cuts. In Sec. \ref{sec:mom} we demonstrate that cutting in longitudinal phase space preserves enough of the angular information of the reaction to reliably extract moments from the data. Finally we give a short summary of our findings in Sec. \ref{sec:summary}.\\ \indent
All presented studies were carried out using toy Monte Carlo (MC) samples generated with the properties stated in the relevant sections. This has the advantage of providing us with clean samples of particular event types with which to evaluate the discussed methods.

\section{\label{sec:lps}Longitudinal phase space plots}
Longitudinal phase space (LPS) plots were introduced by van Hove in 1969 \cite{van_hove_longitudinal_1969,van_hove_final_1969,bialas_longitudinal_1969}. The premise is that at sufficiently high centre-of-mass (CM) energies the phase space is reflected more or less entirely in terms of the longitudinal components of particle momenta. Therefore by neglecting the small transverse components the dimensionality of the phase space is reduced. LPS plots provide means to visualize the reaction kinematics of an $n$-particle final state in an $(n-1)$-dimensional plane. Specifically, for  three particles in the  final state the polar coordinates $q$ and $\omega$ are defined such that
\begin{align}
 q_1 &= \sqrt{\frac{2}{3}}q\sin\left(\omega\right), \\
 q_2 &= \sqrt{\frac{2}{3}}q\sin\left(\omega+\frac{2}{3}\pi\right), \\
 q_3 &= \sqrt{\frac{2}{3}}q\sin\left(\omega+\frac{4}{3}\pi\right), \\
 q &= \left(q_1^2+q_2^2+q_3^2\right)^{\frac{1}{2}},
\end{align}
where $q_i$ denotes the longitudinal momentum component of the {\it{i}}th particle in the CM frame. 
\begin{figure}
 \centering
 \includegraphics[width=8cm]{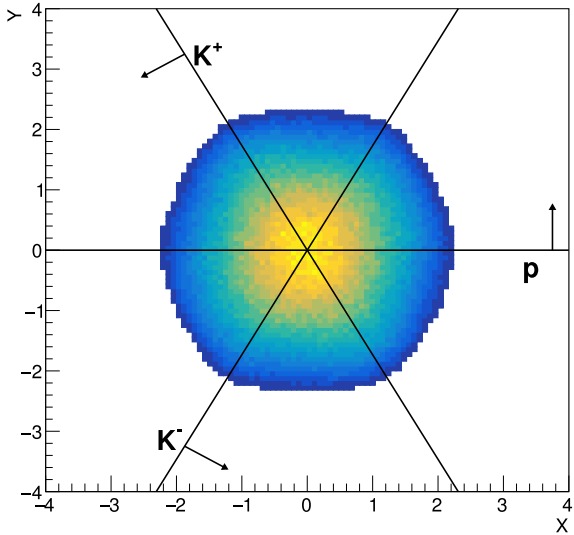}
 \caption{Longitudinal phase space plot of $\gamma p \rightarrow pK^+K^-$ toy events without resonances at $E_{\gamma}=\SI{9}{\GeV}$.}\label{fig:vHPS}
\end{figure}
Figure \ref{fig:vHPS} shows a longitudinal phase space plot of $\gamma p \rightarrow pK^+K^-$ toy events generated according to the phase space distribution at $E_{\gamma}=\SI{9}{\GeV}$. In the plot the coordinates X and Y are defined  by
\begin{align}
 \text{X} &=  q\cos\left(\omega\right),\\
 \text{Y} &=  q\sin\left(\omega\right) \,.
\end{align}
In the plot the solid lines divide the plane into six sectors. The labels and arrows denote a particle and its direction in the CM frame, e.g., all events in the upper half of the plot ($Y>0$) have the outgoing proton going forward as indicated by the arrow. Events in the bottom middle sector have a $K^+$ and $K^-$ going forward and the outgoing proton going backward. At sufficiently high energies and small momentum transfer $t$ this is where one would expect events coming from a meson decaying to a $K^+K^-$  pair.  Events from a baryon decaying to a $K^-p$ final state would be expected in the bottom left sector where the $K^+$ is going forward and the $K^-$ and $p$ are going backward.

\section{\label{sec:cuts}Cuts in longitudinal phase space}
As discussed above reactions with small transverse momentum components, such as {\it{t}}-channel single-Regge exchange, are likely to populate a specific sector of the longitudinal phase space plot at large enough CM energy. Therefore one may simply cut on one of the sectors resulting in an improved ratio of signal to background for the reaction of interest. Here signal and background refer to different processes, e.g., meson or baryon production and decay, which have identical particles in the final state. While this might work well for decays of relatively light particles that are kinematically well contained within one sector, decays of heavy particles occupy a large fraction of phase space and start to leak out into neighboring sectors. As we show here, this effect can be quite sizable. This means that a large part of the signal sample could  be removed from further analysis if one cuts on a sector. Furthermore  the removed part of the signal sample will not be uniformly distributed in angles. Therefore one might introduce unwanted acceptance effects by using such a cut, as demonstrated later. An alternative approach would be to calculate the cut limits in longitudinal phase space, depending on the mass of the decaying particle or resonance in the reaction of interest, such that no signal events are cut. On the other hand one may wish to define cuts to entirely remove the background, if one is unable to model it and in particular if it is interfering with the signal of interest.

\subsection{Mass-dependent cuts}
\begin{figure}[htb]
 \centering
 \includegraphics[width=8cm]{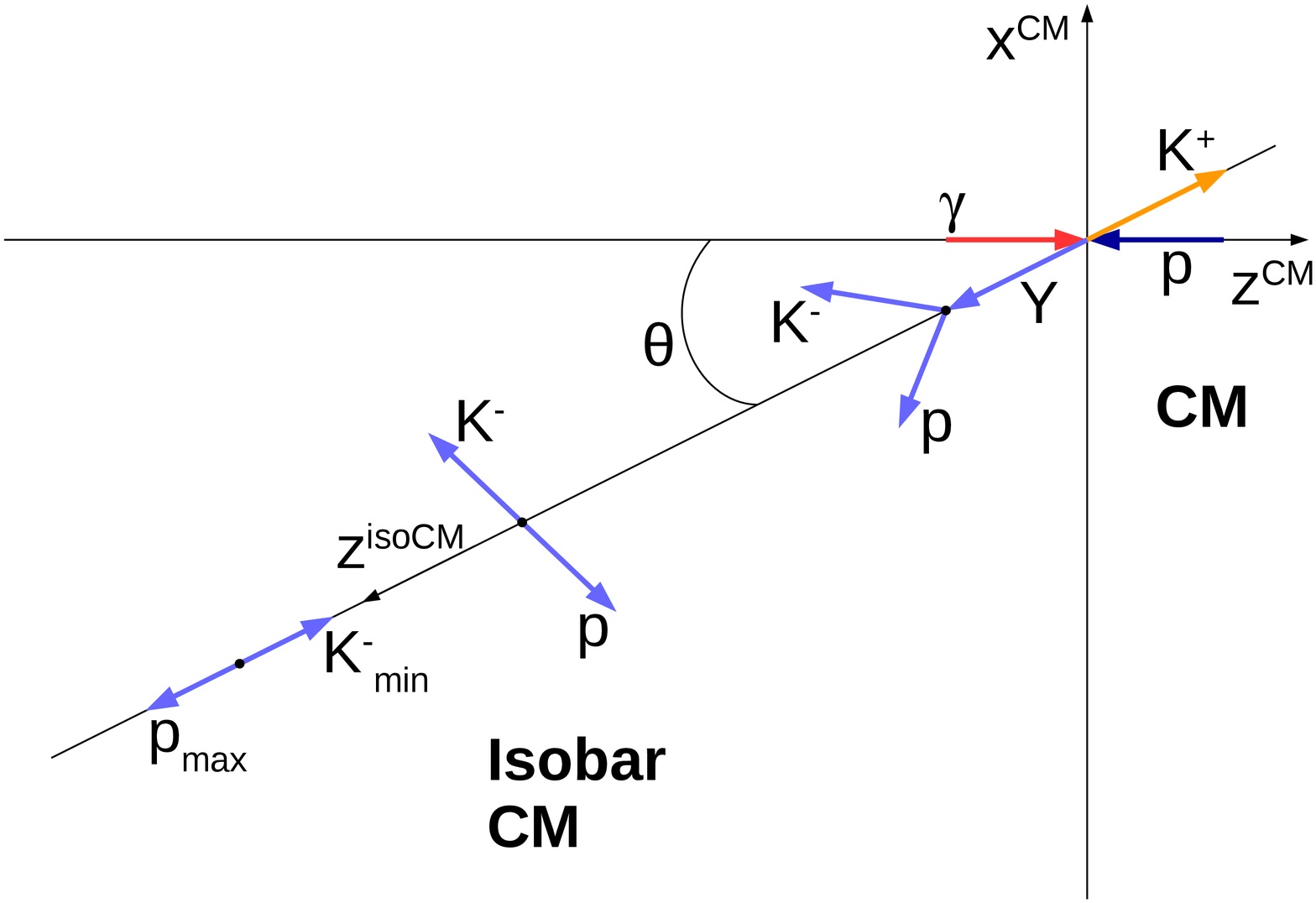}
 \caption{Coordinate systems used to calculate the momentum-dependent cut limits in longitudinal phase space. The directions $p_{\text{max}}$ and $K^{-}_{\text{min}}$ are an example to visualize the case when the proton and kaon have maximal and minimal momentum in the CM frame. \label{fig:coord_system}}
\end{figure}

\begin{figure}[htb]
 \centering
 \includegraphics[width=8cm]{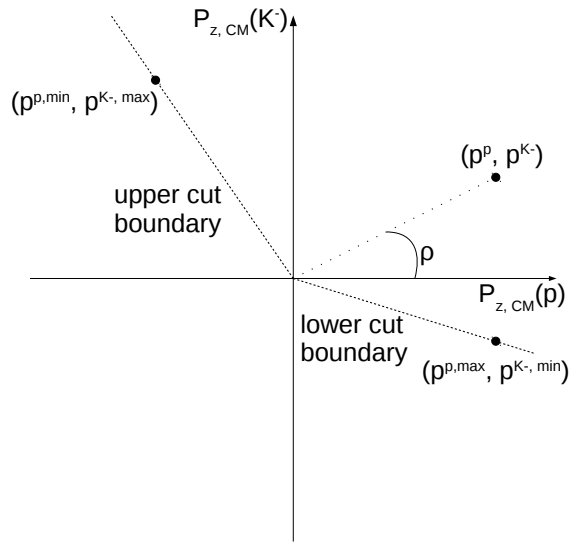}
 \caption{Coordinate system used to evaluate momentum-dependent cut limits. The limits are defined by Eq. \eqref{eq:omega}. The \SI{60}{\degree} sectors of van Hove's LPS plot are remapped on \SI{90}{\degree} sectors for easier cut evaluation.}\label{fig:cutlimit}
\end{figure}

The idea behind a mass-dependent cut in LPS is to ask the question, could two particles in the final state of a reaction have formed an isobar, solely considering their kinematics?\\ \indent
This already implies that the mass-dependent cut is targeted at a specific reaction, i.e., meson or baryon resonance production. We present the formalism for a three particle final state using the reaction $\gamma p\rightarrow K^{+}Y \rightarrow K^{+}K^{-}p$ as an example. A general description is given in the Appendix.\\ \indent
Two reference frames are defined to determine the cut limits. The overall CM frame is defined such that $z^{\text{CM}}$ is along the incoming beam direction and $y^{\text{CM}}$ is perpendicular to $z^{\text{CM}}$ and the direction of the outgoing isobar $Y$, defined by the cross-product of the isobar and the beam photon momenta. $x^{\text{CM}}$ is defined as the cross-product of $y^{\text{CM}}$ and $z^{\text{CM}}$.
The isobar CM frame is defined such that $z^{\text{isoCM}}$ is in the direction of the isobar in the CM frame, $y^{\text{isoCM}}$ is the cross-product of the beam photon in the isobar CM frame and $z^{\text{isoCM}}$, and $x^{\text{isoCM}}=y^{\text{isoCM}}\times z^{\text{isoCM}}$. Figure \ref{fig:coord_system} visualizes both frames.
In the decaying isobar CM frame the $K^-$ and outgoing $p$ are back to back with a fixed momentum given by the standard two-body kinematics for a decay at rest:
\begin{align}
 P =\frac{\left[\left(M_Y^{2}-\left(m_{p}+m_{K^-}\right)^{2}\right)\left(M_Y^{2}-\left(m_{p}-m_{K^-}\right)^{2}\right)\right]^{1/2}}{2M_Y},
\end{align}
with $M_Y$ being the invariant mass of the isobar. That means that if the particle's four-momentum $(p^x, p^y, p^z, E)$ is
\begin{align}
 p^{\text{isoCM}}_{K^-/p} = \left(0,0,\pm P,\sqrt{P^2+m_{K^-/p}}\right) \label{eq:minmax}
\end{align}
in the isobar CM frame, boosting back to the overall CM frame yields the maximal/minimal momentum that the particles can have in the CM frame. In order to evaluate the mass-dependent LPS cut, so as to retain high mass isobar decays, the van Hove angle $\omega$ is replaced with
\begin{align}
 \rho = \arctan{\left(\frac{p^{z,\text{CM}}_{K^-}}{p^{z,\text{CM}}_{p}}\right)} \,,\label{eq:omega}
\end{align}
where $p^{z,\text{CM}}$ indicates that, as in the van Hove plot, only the longitudinal component of the momentum vector in the CM frame is used. Using this definition and the four-vectors calculated in Eq. \eqref{eq:minmax}, boosted into the reaction CM frame, one can calculate the lower and upper cut limits for $\rho$ for a given event as
\begin{align}
 \rho_{\text{low}} & = \arctan{\left(\frac{p^{z,\text{CM}}_{K^-\text{, min}}}{p^{z,\text{CM}}_{p\text{, max}}}\right)}, \\
 \rho_{\text{up}} & = \arctan{\left(\frac{p^{z,\text{CM}}_{K^-\text{, max}}}{p^{z,\text{CM}}_{p\text{, min}}}\right)}.
\end{align}
Note that those cut limits are dependent on the isobar mass. As the isobar mass increases its decay will occupy a larger fraction of phase space, resulting in less rejection of background mechanisms. Figure \ref{fig:cutlimit} shows the isobar mass-dependent cut limits and particle kinematics in the coordinate system used to evaluate the mass-dependent LPS cut.
Reactions that fulfill the condition
\begin{align}
 \rho_\text{low} < \rho < \rho_{\text{up}}\label{eq:rho_limits}
\end{align}
could have formed an isobar produced with momentum parallel (antiparallel) to the beam direction. As in the LPS plots introduced by van Hove this holds only strictly true in the limit of infinite longitudinal momentum, or vanishing transverse momentum. In this case $\theta$ as defined in Fig. \ref{fig:coord_system} is zero. For large longitudinal momenta, i.e., high photon energies and large {\it{t}} slopes, the condition is approximately true. In cases of large momentum transfer the cuts will remove more events of interest and be less effective.

\subsection{Comparison between mass-independent and mass-dependent cuts}
\begin{figure*}[htb]
 \centering
 \includegraphics[width=18cm]{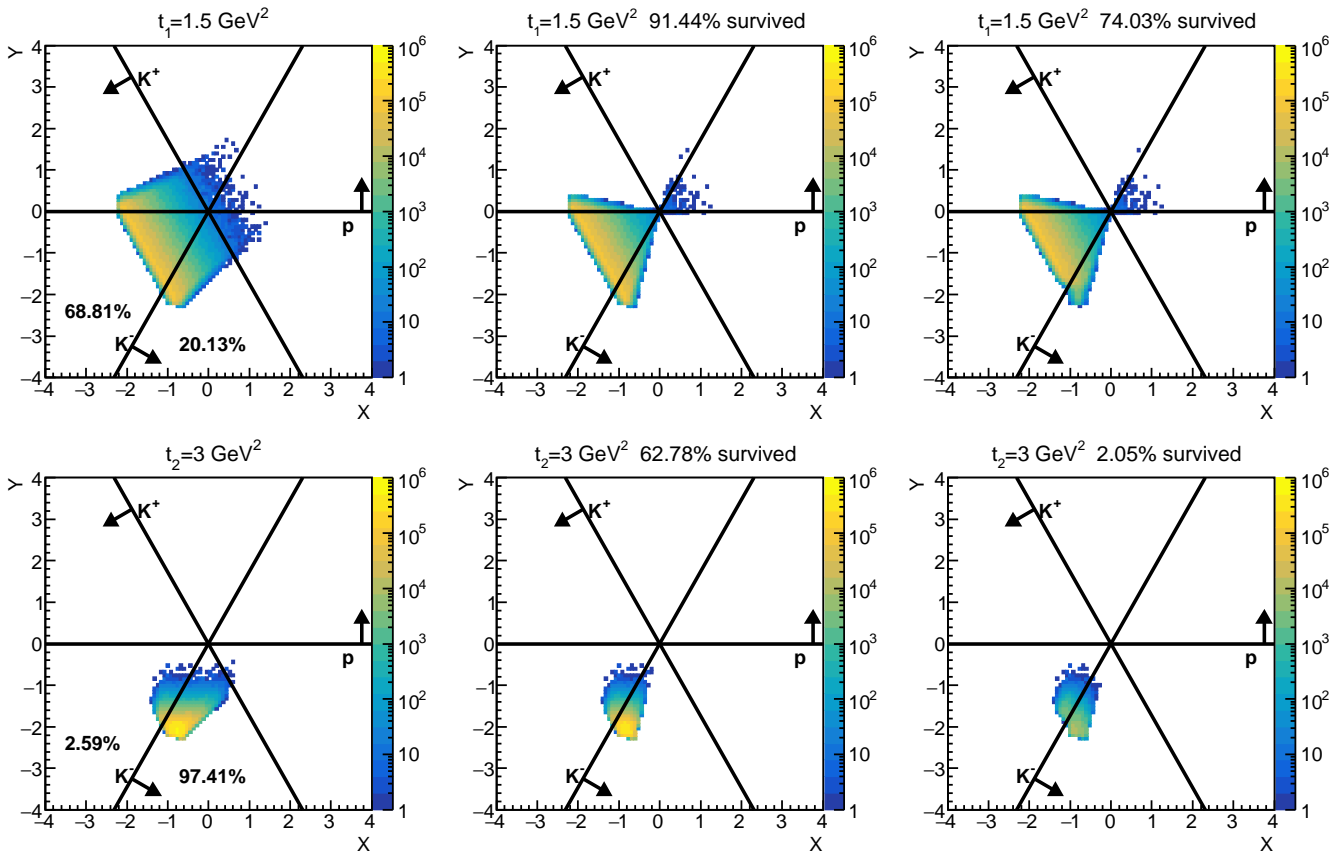}
 \caption{Comparison of the different LPS cuts and the percentages of events surviving the cuts. The plots in the top row originate from baryon isobar events with a {\it{t}} slope of $t_1=\SI{1.5}{\GeV^{2}}$. The plots in the bottom row originate from events with a spectrum of meson isobars with a {\it{t}} slope of $t_2=\SI{3.0}{\GeV^{2}}$. The plots on the left visualize the loss of events by cutting on a specific sector as defined by van Hove. The plots in the middle show the remaining events after applying the baryon isobar cut, and the plots on the right show the remaining events after applying the BNM isobar cut.}\label{fig:vhcomp}
\end{figure*}

\begin{figure*}[htb]
\centering
 \begin{subfigure}[htb]{0.45\textwidth}
 \centering
 \includegraphics[width=8cm]{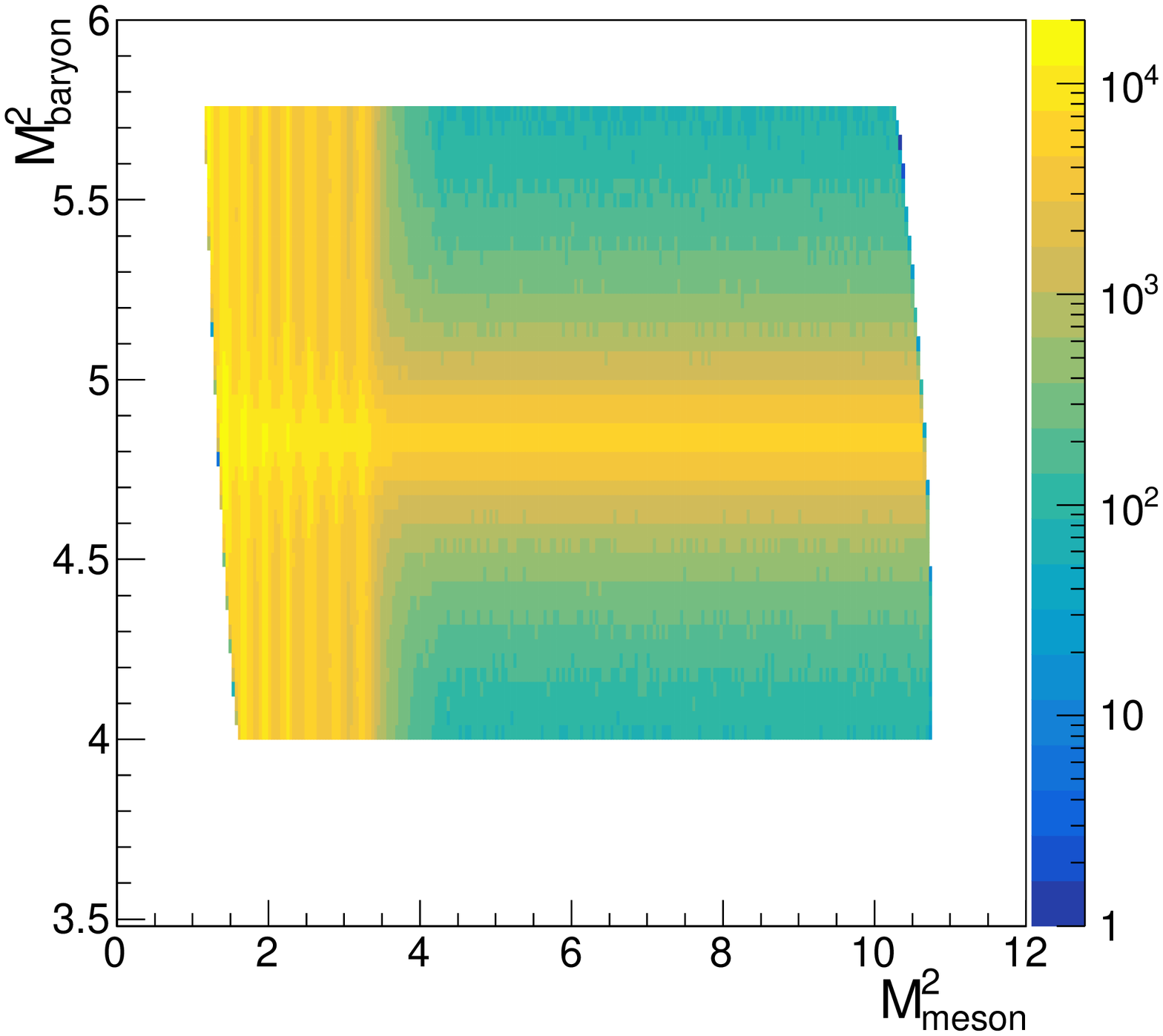}
 \caption{No cut}\label{fig:dalitzNoCut}
\end{subfigure}
\begin{subfigure}[htb]{0.45\textwidth}
 \centering
 \includegraphics[width=8cm]{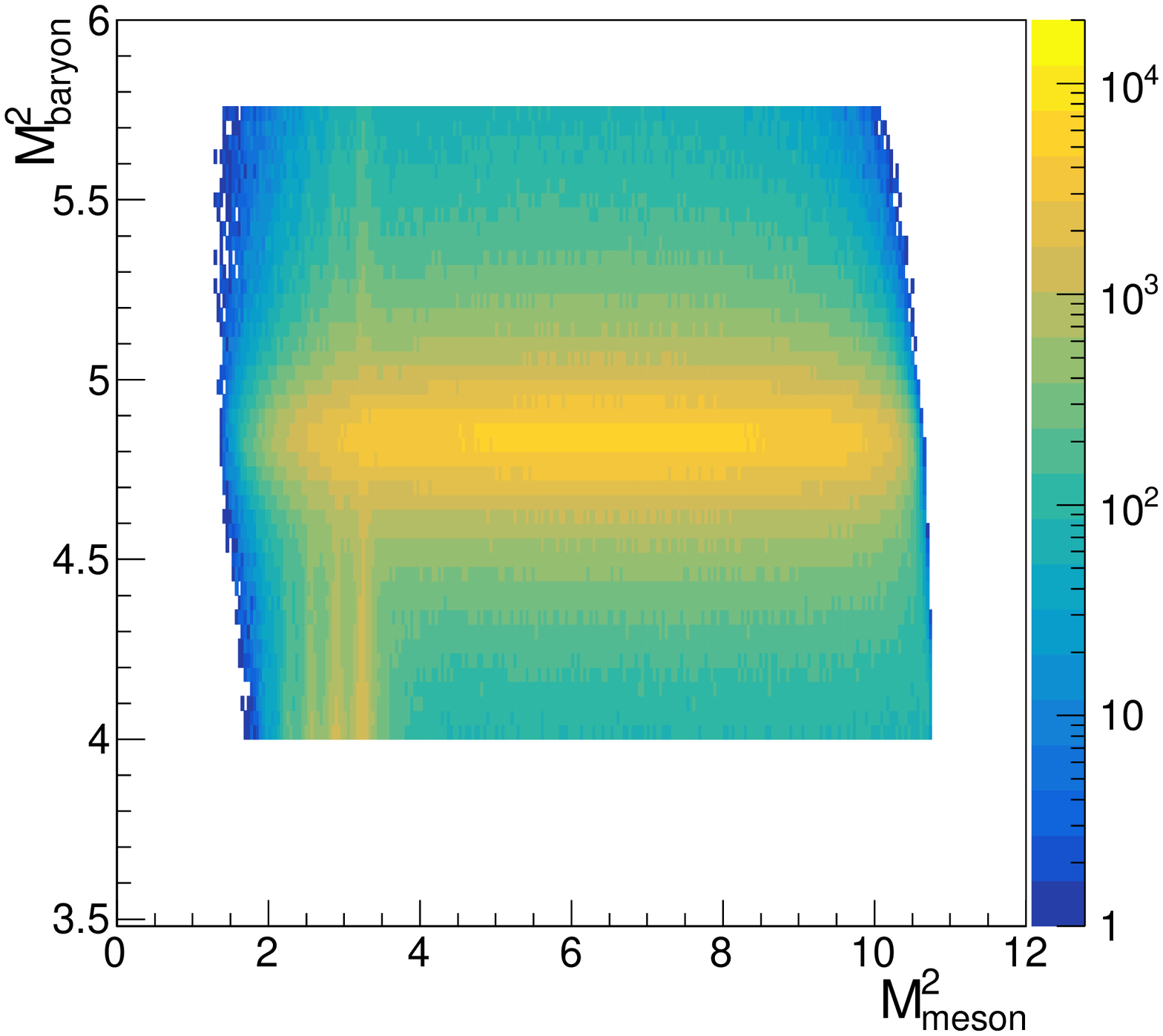}
 \caption{Baryon sector cut}\label{fig:dalitzSectorB}
\end{subfigure}
\begin{subfigure}[htb]{0.45\textwidth}
 \centering
 \includegraphics[width=8cm]{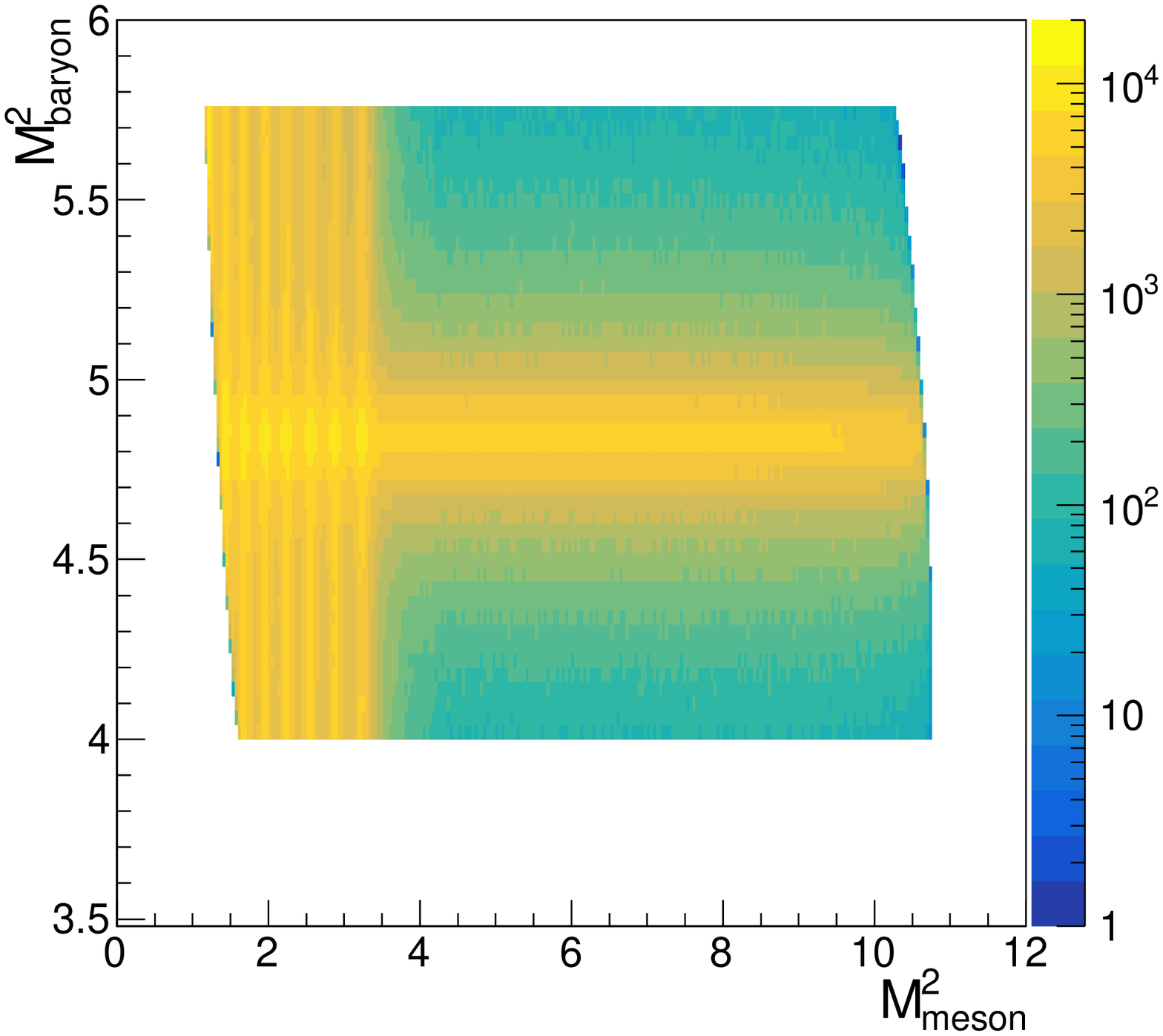}
 \caption{Baryon isobar cut}\label{fig:dalitzbIso}
\end{subfigure}
\begin{subfigure}[htb]{0.45\textwidth}
 \centering
 \includegraphics[width=8cm]{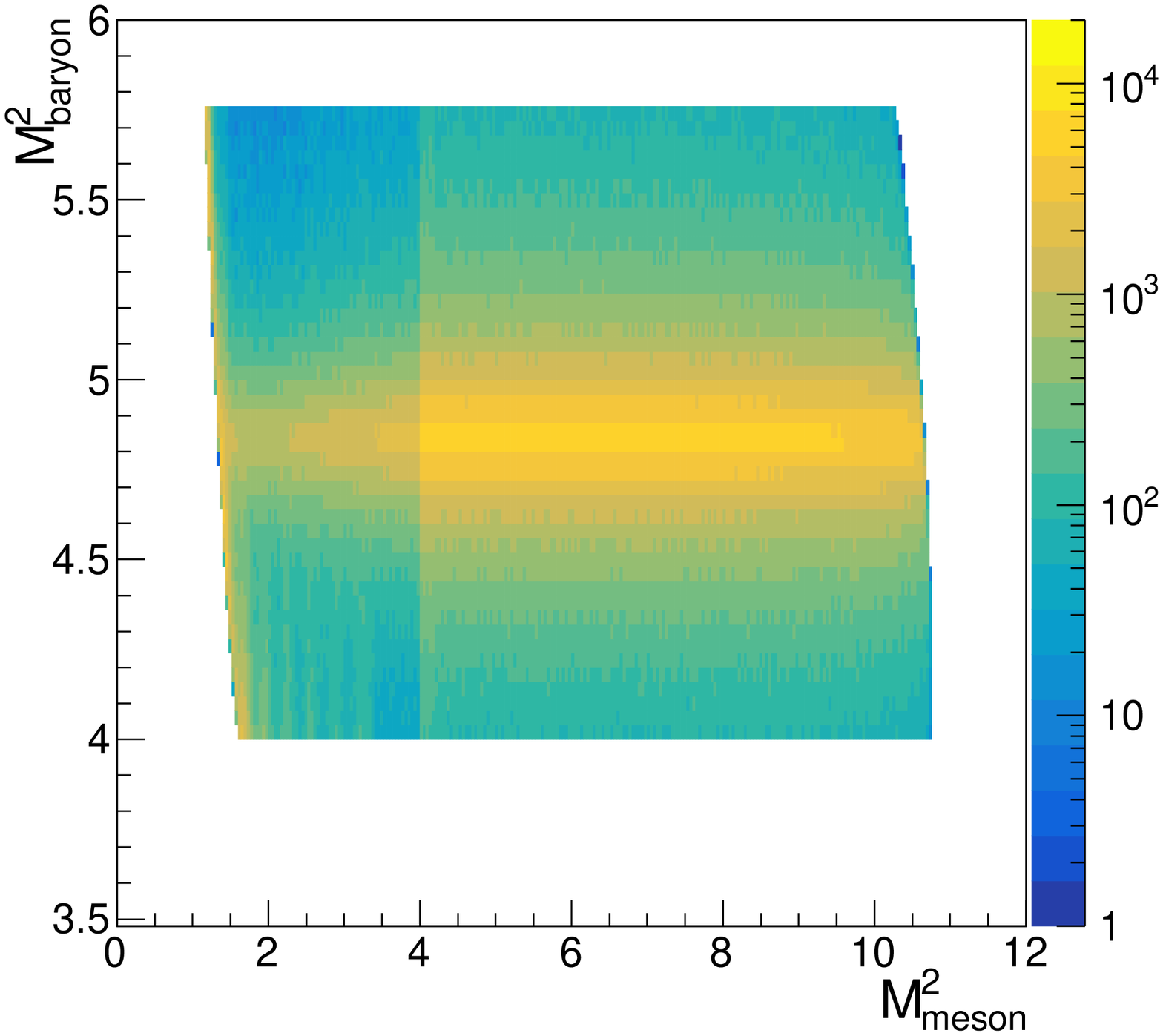}
 \caption{BNM cut}\label{fig:dalitzbIsoNmIso}
\end{subfigure}
\caption{Dalitz plots of the simulated signal and background samples used to evaluate the cut performances. The effects of the studied cuts are nicely visible. The baryon sector cut manages to remove most of the meson background but especially higher mass meson states are not as affected by the cut. The BNM cut performs much better and the meson background is reduced to a minimum.}\label{fig:dalitzComp}
\end{figure*}

\begin{figure}[htb]
 \centering
 \includegraphics[width=8cm]{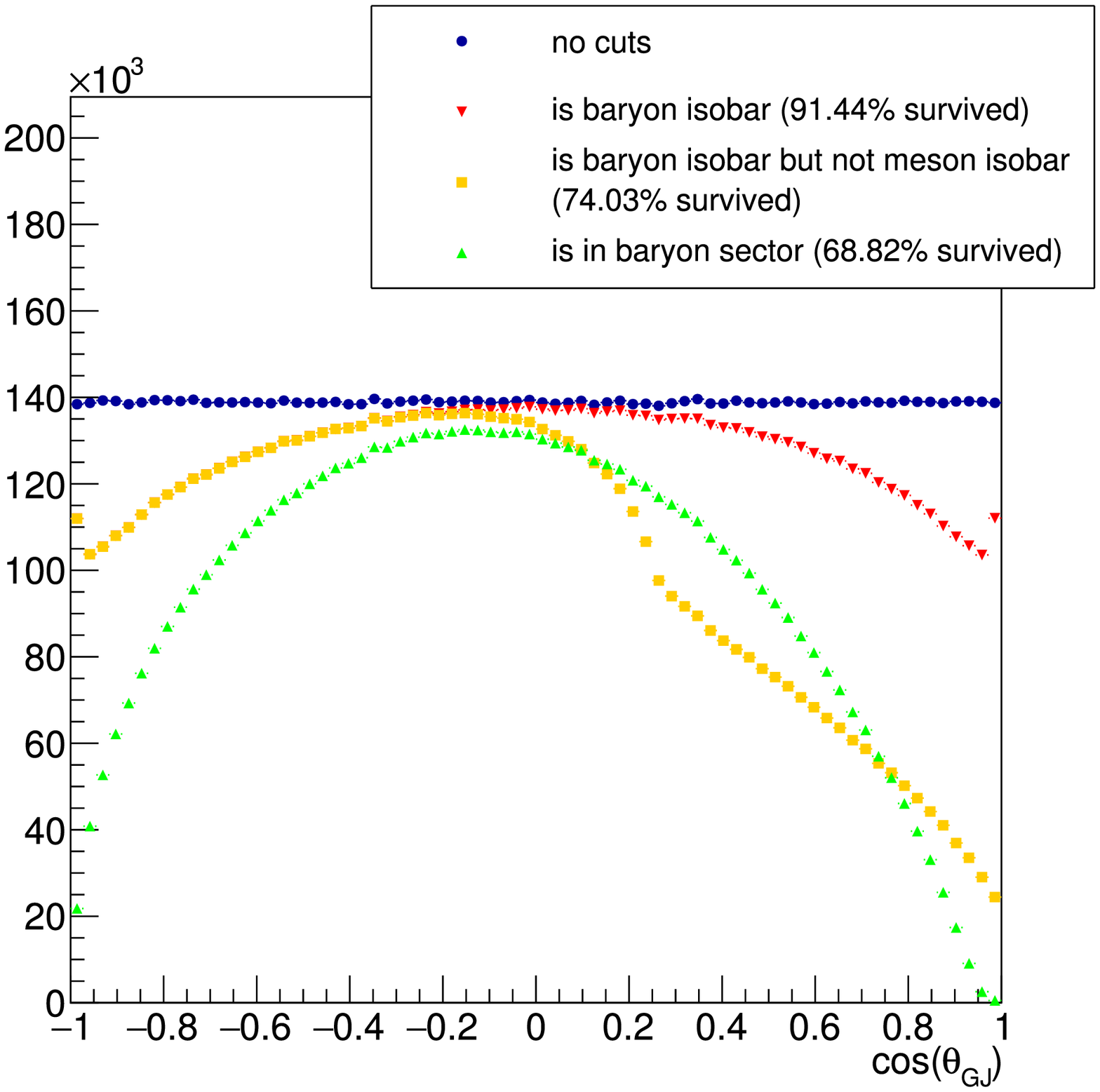}
 \caption{$\cos\left(\theta_{\text{GJ}}\right)$ distributions of events after applying the various cuts studied.}\label{fig:cosTB}
\end{figure}

Neither the van Hove nor the mass-dependent LPS plot distinguishes between baryons or mesons based on their quantum numbers or quark content. Only the longitudinal momenta of the involved final state particles are considered. Therefore it does not matter if one studies the effects of the cuts on selecting mesons or baryons: the method is the same. In the following, standard (van Hove) and mass-dependent cuts were applied to a known sample of generated toy events. The signal reaction used for our study is $\gamma p \rightarrow K^{+}Y(2200)\rightarrow K^{+}K^{-}p$ at $E_{\gamma}=\SI{9}{\GeV}$, where $Y(2200)$ denotes a hyperon with a mass of \SI{2.2}{\GeV}. The width of this generated baryon is \SI{50}{\MeV}. For the event generation a {\it{t}} slope of $t_1=\SI{1.5}{\GeV^{2}}$ was used. In this case $t_1$ is the momentum transfer between the target proton and the produced baryon, and the constant slope is consistent with a single Regge, kaon trajectory exchange. The number was chosen as a lower estimate of what we expect for heavy hyperon production. To study the influence of background reactions, a toy spectrum of nine meson resonances with masses between $1.0$ and \SI{1.8}{\GeV} and widths of \SI{50}{\MeV} were generated. For meson resonance production, the relevant momentum transfer is between the proton target and the proton recoil, and we generated this  background spectrum using a slope of  $t_2=\SI{3.0}{\GeV^{2}}$. $t_{2}$ was chosen according to a measurement of the $\phi(1020)$ {\it{t}} slope \cite{PhysRevC.89.055208}. A cut of $\pm$\SI{0.2}{\GeV} was placed around a $pK^-$-invariant mass of \SI{2.2}{\GeV}, as one might do to study the specific resonance $Y(2200)$.\\ \indent
The effects of three different cuts were studied. The first cut selects all events within the sector of $K^{-}p$ going backwards and $K^{+}$ going forward. This is where baryons coming from {\it{t}}-channel production would be expected. This cut will be referred to as sector cut, as it is equivalent to selecting events in a sector of the van Hove plot. The second cut requires that the $K^{-}$ and the $p$ could have formed an isobar, as discussed previously. This cut will be referred to as baryon isobar cut. The last cut studied for our comparison requires again that the $K^{-}$ and the $p$ could have formed an isobar but also requires that the $K^{+}$ and $K^{-}$ could {\it{not}} come from a meson isobar. This means that $\rho_\text{low}$ and $\rho_\text{up}$ are calculated for a $K^{+}K^{-}$ isobar hypothesis and then the events are cut if their $\rho$ is within these calculated limits. In this case an upper meson isobar mass constraint of \SI{2}{\GeV} is applied so that baryon events, that usually contain a high momentum $K^{+}$ in {\it{t}}-channel production, are less effected by this veto. In real experiments the effects from mesons above this isobar mass limit are expected to be relatively small and it should be tuned to the actual experiment. This cut, specifically rejecting mesons, will be referred to as a baryon-not-meson (BNM) isobar cut.\\ \indent
Figure \ref{fig:vhcomp} shows a comparison of the resulting LPS plots after applying the cuts.
The top and bottom row depict the LPS plots for the baryon and the meson spectrum respectively. The column on the left shows the LPS plots without any cuts applied. As expected, the baryons end up in the bottom left sector of the plot and the mesons in the bottom middle sector. The percentages in the plot show how many events ended up in this particular sector, quantifying how many events would be left after such a cut. The two plots in the centre column show the LPS plot after applying the baryon isobar cut, while the two plots on the right show the LPS plots after applying the BNM isobar cut. It is interesting to see that the baryon isobar cut preserves more than 90\% and the BNM isobar cut around 74\% of the signal events while the sector cut only preserves 68\% of the signal events. This already indicates that mass-dependent cuts are an improvement over simple sector cuts. Furthermore if one looks at the amount of background rejected one can see that the BNM isobar cut rejects about 98\% of the background events while the sector cut rejects around 97.5\%. The BNM isobar cut preserves more signal while having slightly improved background rejection. This can again be seen in Fig. \ref{fig:dalitzComp}.
Four Dalitz plots are shown for the toy sample without a cut (Fig. \ref{fig:dalitzNoCut}) and the studied cuts (Figs. \ref{fig:dalitzSectorB}-\ref{fig:dalitzbIsoNmIso}). Especially the comparison between the sector cut (Fig. \ref{fig:dalitzSectorB}) and the BNM cut (Fig. \ref{fig:dalitzbIsoNmIso}) is interesting. One can clearly see how the BNM cut manages to remove higher mass meson resonances that are less affected by the sector cut since they leak from the meson sector into the baryon sector.\\ \indent
Figure \ref{fig:cosTB} demonstrates another important advantage of the BNM isobar cut over the sector cut. 
It shows the $\cos\left(\theta_{\text{GJ}}\right)$ distribution of the signal events for the four settings of interest. The $\theta_{\text{GJ}}$ denotes the decay angle of the $K^{-}$ in the Gottfried-Jackson (GJ) frame, the $K^{-}p$ rest frame with the {\it{z}} axis pointing along the incoming photon beam \cite{Gottfried1964}. For the toy sample which was generated without any spin or angular momentum in the reaction, one would expect a flat distribution, as seen for the ``no cut'' events in Fig. \ref{fig:cosTB}. If cuts are applied that are not uniform in angles, acceptance effects will be introduced that distort the distribution. One can clearly see how the BNM isobar cut alters the distribution at $\cos\left(\theta_{\text{GJ}}\right)=\pm1$ much less compared to the sector cut. This is important as the $\cos\left(\theta_{\text{GJ}}\right)$ distribution is required in data analyses to extract resonance properties from the data. We show in Sec. \ref{sec:momentResults} how the resulting more favorable acceptance improves the extraction of angular moment parameters from a toy data sample.

\section{\label{sec:mom}Moment extraction using mass-dependent cuts}
After establishing that mass-dependent phase space cuts can be an improvement over cutting on a sector in LPS, we now want to show that it remains possible to extract information from angular distributions after applying such cuts. For that purpose we performed studies using 300 toy Monte Carlo samples and compared the results from samples without any cuts to those with the three cuts outlined in the previous section applied.

\subsection{Monte Carlo setup}
In order to establish if it is possible to extract information from the angular distribution of events that passed the cuts in LPS, we create 300 toy MC samples. Each sample consists of 10 000 events from the reaction $\gamma p \rightarrow K^{+} Y(2200) \rightarrow K^{+}K^{-}p$, where $Y$ denotes a hyperon resonance. For event generation a realistic {\it{t}} slope of $t_1=\SI{1.5}{\GeV^{2}}$ was used. The events were created with an angular distribution in the GJ frame, with the $K^{-}$ momentum chosen to evaluate the decay angles, given by moments of spherical harmonics $\left\langle Y_{LM}\right\rangle$ with $L_{\text{max}}=4$ and $M_{\text{max}}=2$. Spherical harmonic moments, which are related to underlying partial wave amplitudes, provide a general parameterization for particle decays and are often used to summarize experimental results (cf. \cite{PhysRevD.80.072005}). They are represented as
\begin{equation}
\begin{split}
f(\phi,\theta) = \sum_{L=0,M=0}^{L_{\text{max}},M_{\text{max}}} & \left\langle Y_{LM}\right\rangle \cos\left(M\phi\right) \\
& \times\sqrt{\frac{2L+1}{4\pi}\frac{(L-M)!}{(L+M)!}} P_{LM}\left(\cos{\theta}\right)
\end{split}
\end{equation}
where $P_{LM}\left(\cos{\theta}\right)$ denotes the associated Legendre polynomials. For this study the spherical harmonic moments were arbitrarily chosen to be
\begin{align*}
 \left\langle Y_{11}\right\rangle & = 0.3, \\
 \left\langle Y_{20}\right\rangle & = -0.2, \\
 \left\langle Y_{41}\right\rangle & = -0.2, \\
 \left\langle Y_{42}\right\rangle & = 0.1,
\end{align*}
with all other moments set to 0, as we are interested in extracting general information from angular distributions rather than specific physics cases. After generating the events the three cuts were applied and the remaining events were analyzed by an unbinned extended maximum likelihood fit \cite{BARLOW1990496}. We corrected for acceptance effects introduced by the cuts by calculating the normalization integrals through summing phase space Monte Carlo events which had identical cuts applied. This was performed via an extension to the RooFit \cite{Verkerke:2003ir} package.

\subsection{Results}\label{sec:momentResults}
\begin{figure}[htb]
 \centering
 \includegraphics[width=8cm]{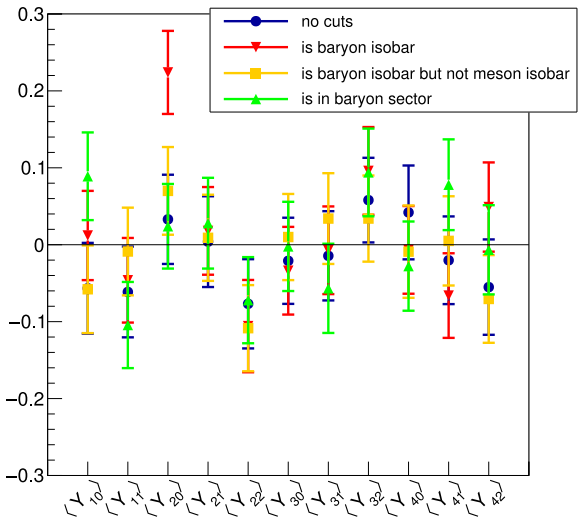}
 \caption{Means of the extracted pull distributions for all the moments considered in the fit.}\label{fig:pullMean}
\end{figure}

\begin{figure}[htb]
 \centering
 \includegraphics[width=8cm]{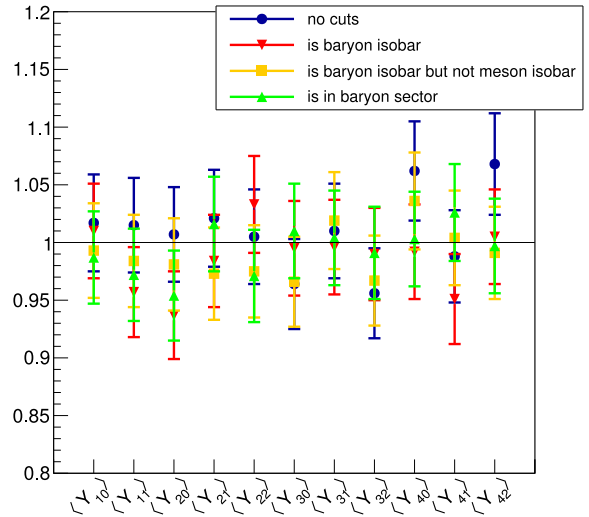}
 \caption{Widths of the extracted pull distributions for all the moments considered in the fit.}\label{fig:pullWidth}
\end{figure}

\begin{figure*}[htb]
\centering
 \begin{subfigure}[htb]{0.45\textwidth}
 \centering
 \includegraphics[width=8cm]{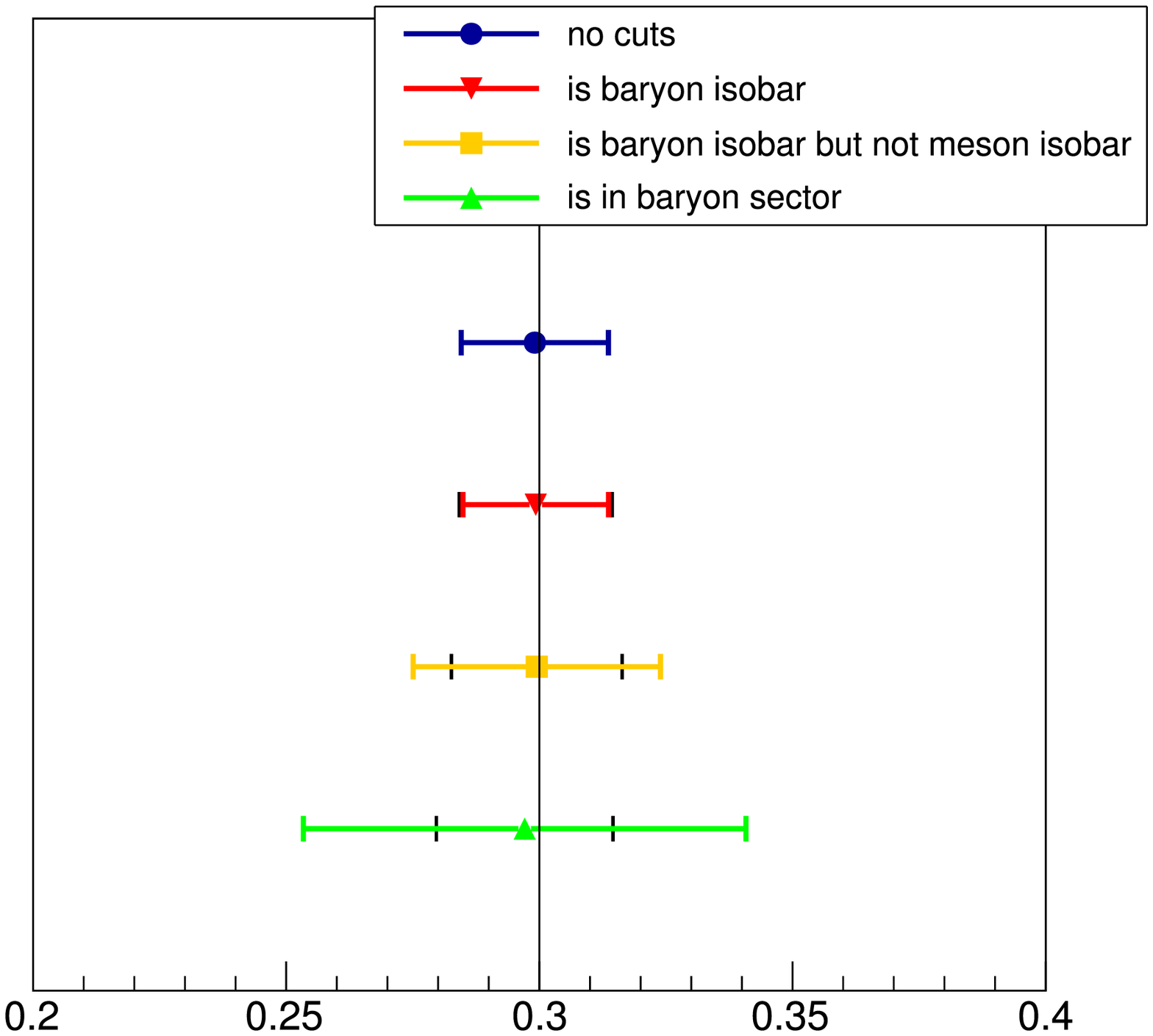}
 \caption{$\langle Y_{11}\rangle$ moment}\label{fig:comp11}
\end{subfigure}
\begin{subfigure}[htb]{0.45\textwidth}
 \centering
 \includegraphics[width=8cm]{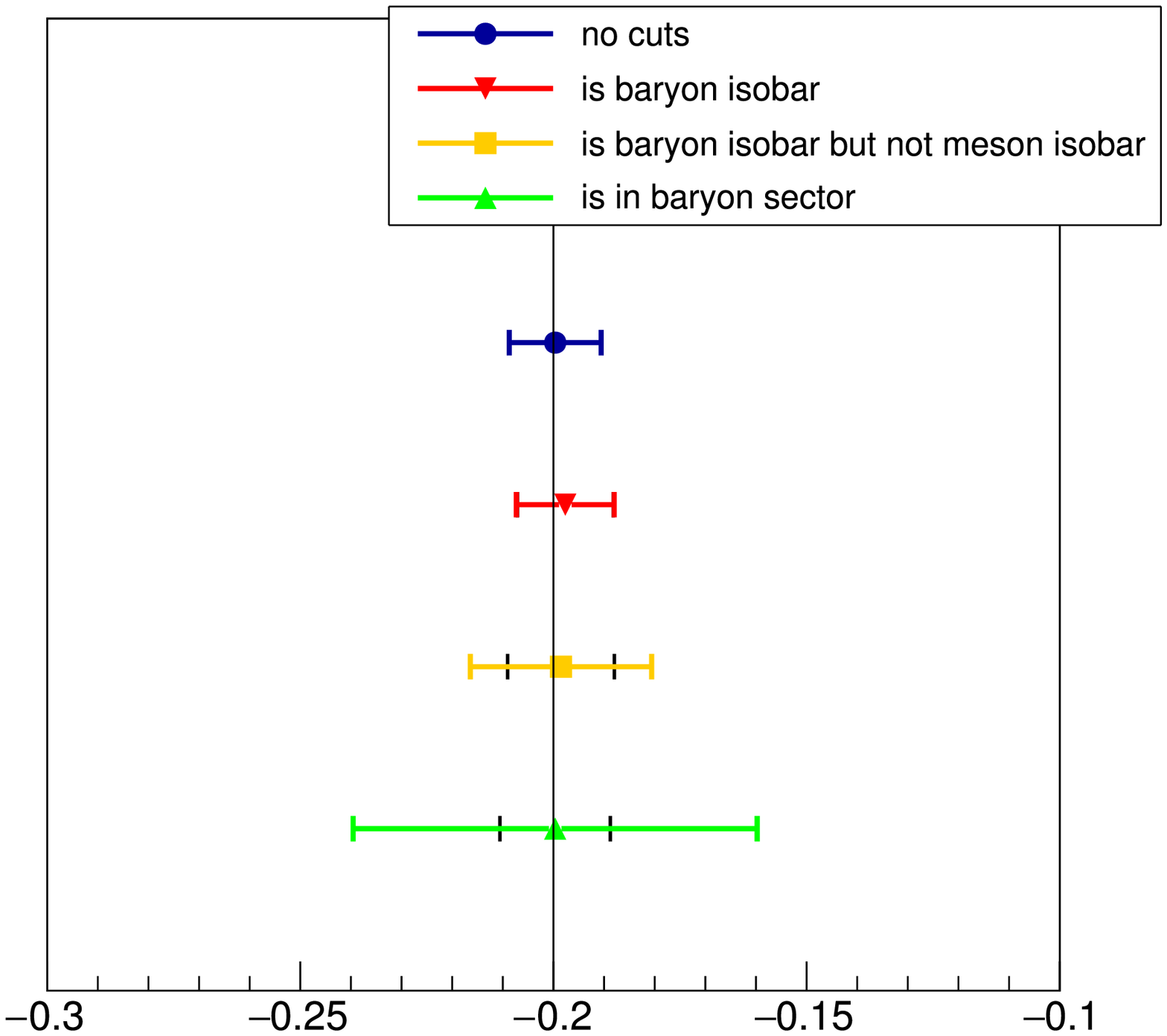}
 \caption{$\langle Y_{20}\rangle$ moment}\label{fig:comp20}
\end{subfigure}
\begin{subfigure}[htb]{0.45\textwidth}
 \centering
 \includegraphics[width=8cm]{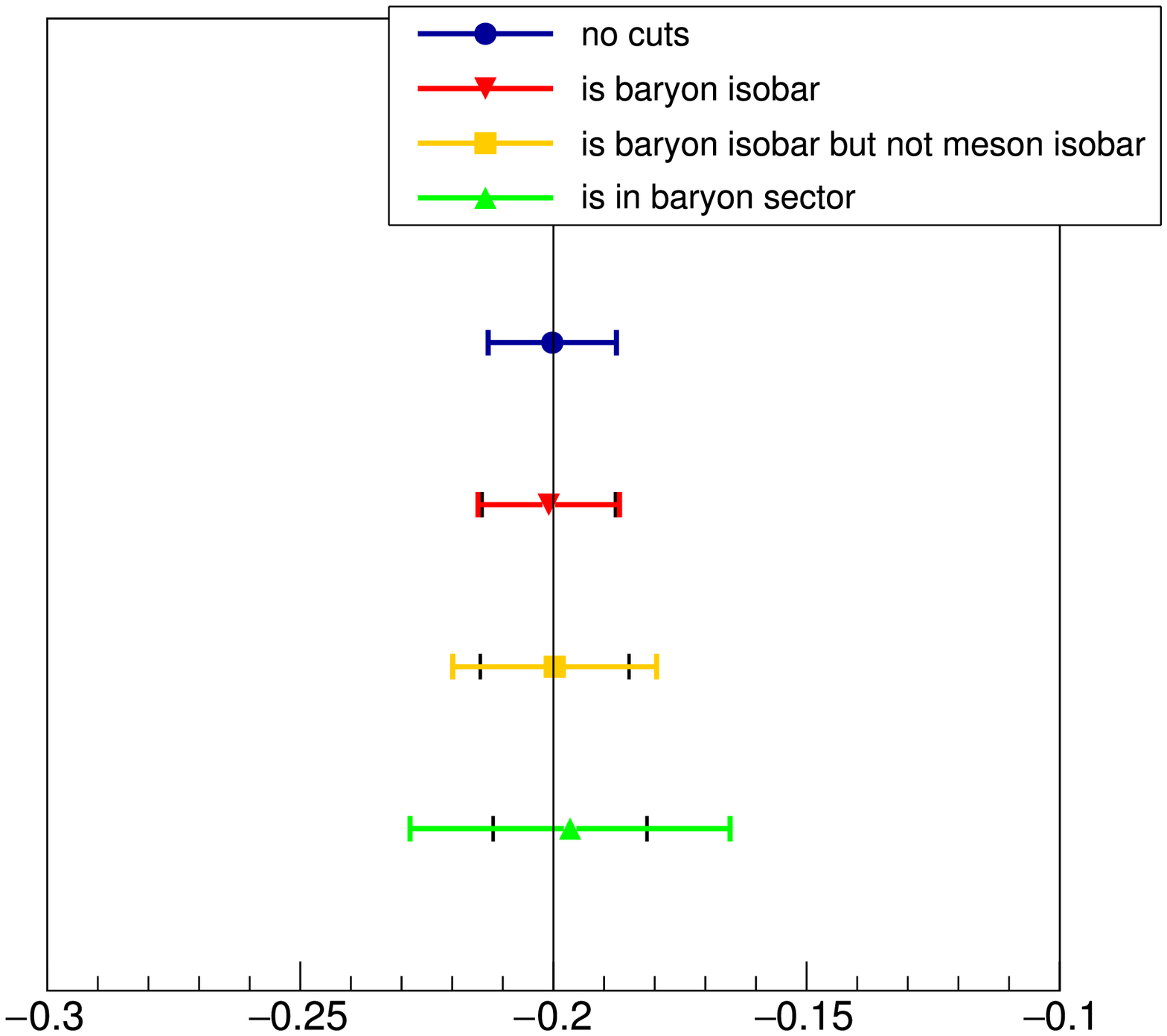}
 \caption{$\langle Y_{41}\rangle$ moment}\label{fig:comp41}
\end{subfigure}
\begin{subfigure}[htb]{0.45\textwidth}
 \centering
 \includegraphics[width=8cm]{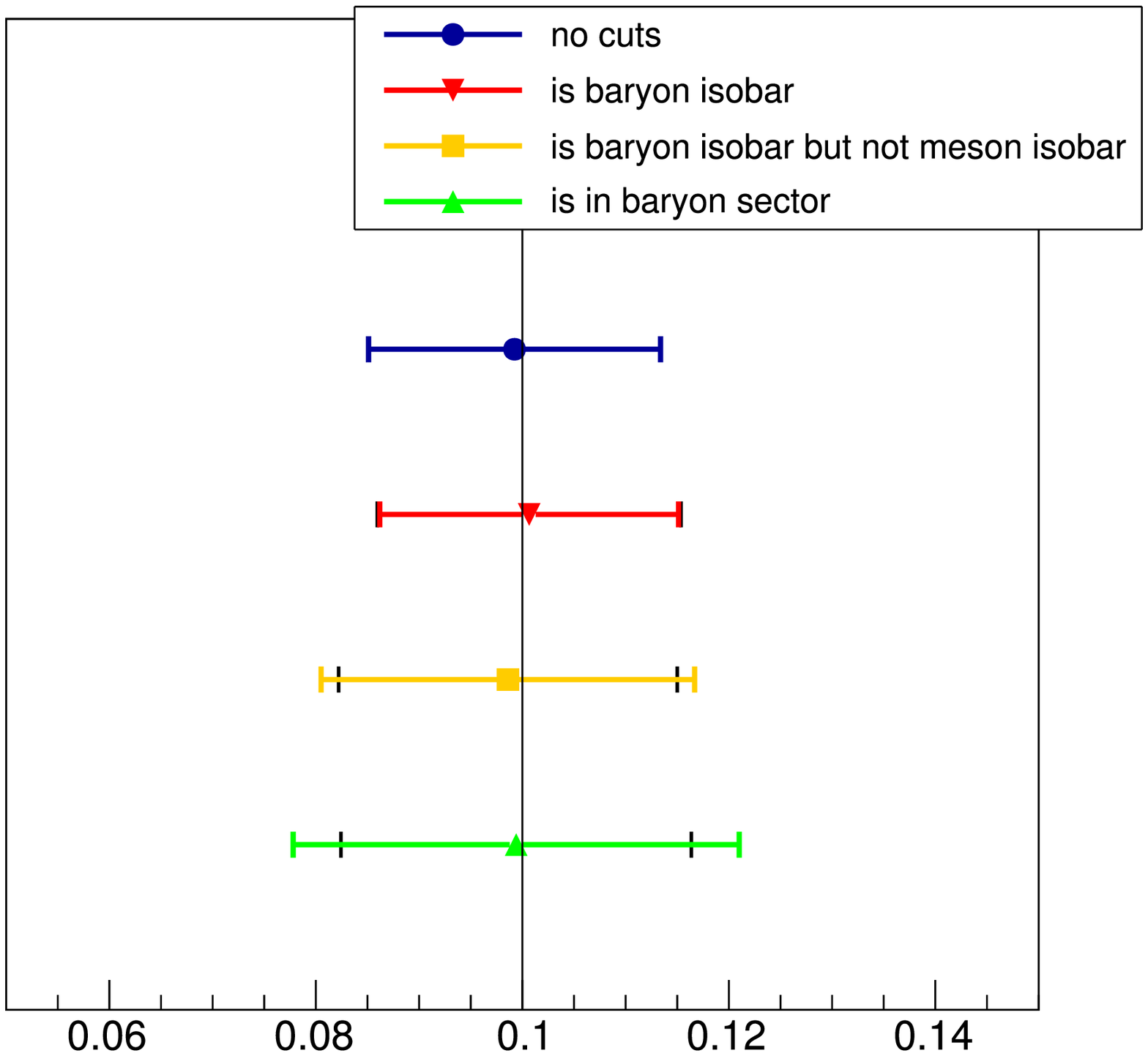}
 \caption{$\langle Y_{42}\rangle$ moment}\label{fig:comp42}
\end{subfigure}
\caption{Comparison of the mean value and standard deviation of nonzero moments. The black error bars in the plots indicate the expected error just taking statistical differences between the datasets into account.}\label{fig:compMom}
\end{figure*}

A common measure to quantify the performance of a fit method are pull distributions. The pull of a measured quantity can be calculated as 
\begin{align}
 \text{pull} = \frac{\text{measured value} - \text{true value}}{\text{error on measured value}}.
\end{align}
If the results are unbiased and the associated uncertainties are well estimated, one would expect a Gaussian pull distribution with a mean of 0 and a width of 1. The mean values and widths for all extracted moments of the four tested settings are shown in Figs. \ref{fig:pullMean} and \ref{fig:pullWidth}.
\\ \indent
One can see that in all four cases the mean values were 0 and the widths were 1 within uncertainties, which were extracted from a Gaussian fit to the pull distributions. This is an important result, showing that all three cuts preserve sufficient information to extract the decay angle distribution characteristics from the data. Even after the cut on the baryon sector, the most restrictive of the four tested cuts, it was possible to extract the information reliably for all zero and nonzero moments. That shows that one is not limited in extraction of small contributions in the angular distribution by these cuts but only by the available statistics.\\ \indent
Pull distributions show if the information extracted from the fits are unbiased and if the errors are determined correctly. In order to also assess the statistical precision of the extracted results it is necessary to compare the extracted parameter values directly. Figure \ref{fig:compMom} shows a direct comparison of the nonzero moments for all four settings.
Here the values and their associated uncertainties were extracted as the mean value and width of a Gaussian fit to the distribution of extracted $\langle Y_{LM}\rangle$ moments. The true values imposed on the toy data sets are indicated by the vertical lines. The black error bars indicate the expected errors of the moments based on the assumption that they originate purely from counting statistics and are growing with $\sqrt{N}$. As expected from the pull distributions, all 16 data points agree with their true value within error bars. Smaller parameter uncertainties are more desirable as they indicate a more precise overall measurement. As we already demonstrated, no systematic effects are induced by these cuts. Unsurprisingly, the distribution without any cuts applied performed best. These datasets have the best statistics and no acceptance effects introduced by the cuts. Comparing the error bars of the moments extracted from the dataset with the baryon sector cut applied to the moments extracted from the dataset with the BNM isobar cut applied, one notices a significant difference. Although the statistics of the events surviving the cuts are comparable, the BNM isobar cut performed much better and resulted in smaller uncertainties for the extracted moments.

\section{\label{sec:summary}Summary}
In this paper we have presented the effects of cuts in longitudinal phase space. This provides a systematic approach for analyzing reactions which may have different processes contributing and quantum mechanically interfering by removing kinematic regions where these processes overlap. We compared the efficiencies of cuts on straight sectors in LPS to a mass-dependent approach. We have shown that a combination of mass-dependent cuts (BNM) can perform better in preserving signal as well as reducing background events than a simple cut on a sector in LPS plots. This innovative new style of background rejection appears very promising and its use in upcoming GlueX and also CLAS12 analyses, as it is equally applicable to {\it{t}}- and {\it{u}}-channel processes in electroproduction, should be considered.\\ \indent
We have also shown that cuts in LPS preserve sufficient information to extract decay parameters from the angular distributions of the particles. Here also mass-dependent cuts in LPS seem to be capable of outperforming cuts on a sector in the LPS plot.\\ \indent
It is important to note that these cuts and their effectiveness are very much dependent on the masses of the involved particles and the {\it{t}} slopes of the involved reactions. Therefore we recommend they should all be studied for each reaction channel; if other aspects of the analysis are correctly accounted for, then any cut method should provide consistent final results.

\section*{\label{sec:acknowledgments}Acknowledgments}
This work was supported by the Scottish Universities Physics Alliance (SUPA),
the United Kingdom's Science and Technology Facilities Council under Grant No. ST/P004458/1, the US Department of Energy, Office of Science, Office of Nuclear Physics under Contracts No. DE-AC05-06OR2317, No. DE-AC05-06OR23177, and No. DE-FG02-87ER40365, and the U.S. National Science Foundation under Grants No. PHY-1415459, No. PHY-1205019, and  No. PHY-1513524. 

\section*{\label{sec:appendix}Appendix: General formalism for mass-dependent cuts in LPS}
The formalism presented in Sec. \ref{sec:cuts} uses the reaction $\gamma p \rightarrow K^+Y\rightarrow K^+K^-p$ as an example. Here we want to present the same formalism for the general case of $\gamma p \rightarrow X R\rightarrow X Y Z$. Recall that the mass-dependent LPS cut purely considers the kinematics of a reaction. Therefore $R$ can be either a meson or a baryon resonance and the equations are equally valid.\\ \indent
The coordinate systems are defined exactly as in Sec. \ref{sec:cuts}. The overall CM frame is defined such that $z^{\text{CM}}$ is along the incoming beam direction and $y^{\text{CM}}$ is perpendicular to $z^{\text{CM}}$ and the direction of the outgoing isobar $Y$, defined by the cross-product of the isobar and the beam photon momenta. $x^{\text{CM}}$ is defined as the cross-product of $y^{\text{CM}}$ and $z^{\text{CM}}$.
The isobar CM frame is defined such that $z^{\text{isoCM}}$ is in the direction of the isobar in the CM frame, $y^{\text{isoCM}}$ is the cross-product of the beam photon in the isobar CM frame and $z^{\text{isoCM}}$ and $x^{\text{isoCM}}=y^{\text{isoCM}}\times z^{\text{isoCM}}$.\\ \indent
The momentum of $Y$ and $Z$ in the rest frame of $R$ is given by
\begin{align}
 P =\frac{\left[\left(M_R^{2}-\left(m_{Y}+m_{Z}\right)^{2}\right)\left(M_R^{2}-\left(m_{Y}-m_{Z}\right)^{2}\right)\right]^{1/2}}{2M_R} \,,
\end{align}
with $M_R$ being the invariant mass of the intermediate resonance. That means that if the particle's four-momenta $(p^x, p^y, p^z, E)$, which must be back to back, are
\begin{align}
 p_Y^{\text{isoCM}} &= \left(0,0,\pm P,\sqrt{P^2+m_Y}\right), \\
 p_Z^{\text{isoCM}} &= \left(0,0,\mp P,\sqrt{P^2+m_Z}\right) \, ,
\end{align}
then boosting back to the overall CM frame yields the maximum ($+P$ in rest frame of $R$) and minimum ($-P$ in rest frame of $R$) momenta that the particles can have in the CM frame  (denoted by $p^{z,\text{CM}}_{Y\text{, min}}$, cf. Fig. \ref{fig:coord_system}). To evaluate the cut limits of the mass-dependent LPS cut, the angle $\rho$ is defined as
\begin{align}
 \rho = \arctan{\left(\frac{p^{z,\text{CM}}_{Y}}{p^{z,\text{CM}}_{Z}}\right)} \,,
\end{align}
where $p_{Y/Z}^{z,\text{CM}}$ denotes the longitudinal momentum component of the particle $Y$ or $Z$ in the overall CM frame. Using the maximum and minimum momenta, the cut limits are then defined as
\begin{align}
 \rho_{\text{low}} & = \arctan{\left(\frac{p^{z,\text{CM}}_{Y\text{, min}}}{p^{z,\text{CM}}_{Z\text{, max}}}\right)}, \\
 \rho_{\text{up}} & = \arctan{\left(\frac{p^{z,\text{CM}}_{Y\text{, max}}}{p^{z,\text{CM}}_{Z\text{, min}}}\right)}.
\end{align}

%

\begin{thebibliography}{10}%
\makeatletter
\providecommand \@ifxundefined [1]{%
 \@ifx{#1\undefined}
}%
\providecommand \@ifnum [1]{%
 \ifnum #1\expandafter \@firstoftwo
 \else \expandafter \@secondoftwo
 \fi
}%
\providecommand \@ifx [1]{%
 \ifx #1\expandafter \@firstoftwo
 \else \expandafter \@secondoftwo
 \fi
}%
\providecommand \natexlab [1]{#1}%
\providecommand \enquote  [1]{``#1''}%
\providecommand \bibnamefont  [1]{#1}%
\providecommand \bibfnamefont [1]{#1}%
\providecommand \citenamefont [1]{#1}%
\providecommand \href@noop [0]{\@secondoftwo}%
\providecommand \href [0]{\begingroup \@sanitize@url \@href}%
\providecommand \@href[1]{\@@startlink{#1}\@@href}%
\providecommand \@@href[1]{\endgroup#1\@@endlink}%
\providecommand \@sanitize@url [0]{\catcode `\\12\catcode `\$12\catcode
  `\&12\catcode `\#12\catcode `\^12\catcode `\_12\catcode `\%12\relax}%
\providecommand \@@startlink[1]{}%
\providecommand \@@endlink[0]{}%
\providecommand \url  [0]{\begingroup\@sanitize@url \@url }%
\providecommand \@url [1]{\endgroup\@href {#1}{\urlprefix }}%
\providecommand \urlprefix  [0]{URL }%
\providecommand \Eprint [0]{\href }%
\providecommand \doibase [0]{http://dx.doi.org/}%
\providecommand \selectlanguage [0]{\@gobble}%
\providecommand \bibinfo  [0]{\@secondoftwo}%
\providecommand \bibfield  [0]{\@secondoftwo}%
\providecommand \translation [1]{[#1]}%
\providecommand \BibitemOpen [0]{}%
\providecommand \bibitemStop [0]{}%
\providecommand \bibitemNoStop [0]{.\EOS\space}%
\providecommand \EOS [0]{\spacefactor3000\relax}%
\providecommand \BibitemShut  [1]{\csname bibitem#1\endcsname}%
\let\auto@bib@innerbib\@empty
\bibitem [{\citenamefont {Shi}\ \emph {et~al.}(2015)\citenamefont {Shi},
  \citenamefont {Danilkin}, \citenamefont {Fern\'andez-Ram\'irez},
  \citenamefont {Mathieu}, \citenamefont {Pennington}, \citenamefont {Schott},\
  and\ \citenamefont {Szczepaniak}}]{shi_double-regge_2015}%
  \BibitemOpen
  \bibfield  {author} {\bibinfo {author} {\bibfnamefont {M.}~\bibnamefont
  {Shi}}, \bibinfo {author} {\bibfnamefont {I.}~\bibnamefont {Danilkin}},
  \bibinfo {author} {\bibfnamefont {C.}~\bibnamefont {Fern\'andez-Ram\'irez}},
  \bibinfo {author} {\bibfnamefont {V.}~\bibnamefont {Mathieu}}, \bibinfo
  {author} {\bibfnamefont {M. R.}~\bibnamefont {Pennington}}, \bibinfo {author}
  {\bibfnamefont {D.}~\bibnamefont {Schott}}, \ and\ \bibinfo {author}
  {\bibfnamefont {A. P.}~\bibnamefont {Szczepaniak}},\ }\href {\doibase
  10.1103/PhysRevD.91.034007} {\bibfield  {journal} {\bibinfo  {journal} {Phys.
  Rev. D}\ }\textbf {\bibinfo {volume} {91}},\ \bibinfo {pages} {034007}
  (\bibinfo {year} {2015})}\BibitemShut {NoStop}%
\bibitem [{\citenamefont {Glazier}(2016)}]{Glazier:MesonEx}%
  \BibitemOpen
  \bibfield  {author} {\bibinfo {author} {\bibfnamefont {D.~I.}\ \bibnamefont
  {Glazier}},\ }\bibfield  {journal}
  {\href {\doibase 10.1063/1.4949451} {\bibinfo  {journal} {AIP Conference Proceedings}}\ }\textbf {\bibinfo
  {volume} {1735}},\ \bibinfo {pages} {070003} (\bibinfo {year}
  {2016})\BibitemShut {NoStop}%
\bibitem [{\citenamefont
  {Van~Hove}(1969{\natexlab{a}})}]{van_hove_longitudinal_1969}%
  \BibitemOpen
  \bibfield  {author} {\bibinfo {author} {\bibfnamefont {L.}~\bibnamefont
  {Van~Hove}},\ }\href {\doibase 10.1016/0550-3213(69)90133-3} {\bibfield
  {journal} {\bibinfo  {journal} {Nucl. Phys. B}\ }\textbf {\bibinfo {volume}
  {9}},\ \bibinfo {pages} {331} (\bibinfo {year}
  {1969}{\natexlab{a}})}\BibitemShut {NoStop}%
\bibitem [{\citenamefont {Van~Hove}(1969{\natexlab{b}})}]{van_hove_final_1969}%
  \BibitemOpen
  \bibfield  {author} {\bibinfo {author} {\bibfnamefont {L.}~\bibnamefont
  {Van~Hove}},\ }\href {\doibase 10.1016/0370-2693(69)90343-8} {\bibfield
  {journal} {\bibinfo  {journal} {Phys. Lett. B}\ }\textbf {\bibinfo {volume}
  {28}},\ \bibinfo {pages} {429} (\bibinfo {year}
  {1969}{\natexlab{b}})}\BibitemShut {NoStop}%
\bibitem [{\citenamefont {Bia\l{}as}\ \emph {et~al.}(1969)\citenamefont
  {Bia\l{}as}, \citenamefont {Eskreys}, \citenamefont {Kittel}, \citenamefont
  {Pokorski}, \citenamefont {Tuominiemi},\ and\ \citenamefont
  {Van~Hove}}]{bialas_longitudinal_1969}%
  \BibitemOpen
  \bibfield  {author} {\bibinfo {author} {\bibfnamefont {A.}~\bibnamefont
  {Bia\l{}as}}, \bibinfo {author} {\bibfnamefont {A.}~\bibnamefont {Eskreys}},
  \bibinfo {author} {\bibfnamefont {W.}~\bibnamefont {Kittel}}, \bibinfo
  {author} {\bibfnamefont {S.}~\bibnamefont {Pokorski}}, \bibinfo {author}
  {\bibfnamefont {J.~K.}\ \bibnamefont {Tuominiemi}}, \ and\ \bibinfo {author}
  {\bibfnamefont {L.}~\bibnamefont {Van~Hove}},\ }\href {\doibase
  10.1016/0550-3213(69)90096-0} {\bibfield  {journal} {\bibinfo  {journal}
  {Nucl. Phys. B}\ }\textbf {\bibinfo {volume} {11}},\ \bibinfo {pages} {479}
  (\bibinfo {year} {1969})}\BibitemShut {NoStop}%
\bibitem [{\citenamefont {Dey}\ \emph {et~al.}(2014)\citenamefont {Dey} \emph
  {et~al.}}]{PhysRevC.89.055208}%
  \BibitemOpen
  \bibfield  {author} {\bibinfo {author} {\bibfnamefont {B.}~\bibnamefont
  {Dey}} \emph {et~al.} (\bibinfo {collaboration} {CLAS Collaboration}),\
  }\href {\doibase 10.1103/PhysRevC.89.055208} {\bibfield  {journal} {\bibinfo
  {journal} {Phys. Rev. C}\ }\textbf {\bibinfo {volume} {89}},\ \bibinfo
  {pages} {055208} (\bibinfo {year} {2014})}\BibitemShut {NoStop}%
\bibitem [{\citenamefont {Gottfried}\ and\ \citenamefont
  {Jackson}(1964)}]{Gottfried1964}%
  \BibitemOpen
  \bibfield  {author} {\bibinfo {author} {\bibfnamefont {K.}~\bibnamefont
  {Gottfried}}\ and\ \bibinfo {author} {\bibfnamefont {J.~D.}\ \bibnamefont
  {Jackson}},\ }\href {\doibase 10.1007/BF02750195} {\bibfield  {journal}
  {\bibinfo  {journal} {Nuovo Cimento}\ }\textbf {\bibinfo {volume} {33}},\
  \bibinfo {pages} {309} (\bibinfo {year} {1964})}\BibitemShut {NoStop}%
\bibitem [{\citenamefont {Battaglieri}\ \emph {et~al.}(2009)\citenamefont
  {Battaglieri} \emph {et~al.}}]{PhysRevD.80.072005}%
  \BibitemOpen
  \bibfield  {author} {\bibinfo {author} {\bibfnamefont {M.}~\bibnamefont
  {Battaglieri}} \emph {et~al.} (\bibinfo {collaboration} {CLAS
  Collaboration}),\ }\href {\doibase 10.1103/PhysRevD.80.072005} {\bibfield
  {journal} {\bibinfo  {journal} {Phys. Rev. D}\ }\textbf {\bibinfo {volume}
  {80}},\ \bibinfo {pages} {072005} (\bibinfo {year} {2009})}\BibitemShut
  {NoStop}%
\bibitem [{\citenamefont {Barlow}(1990)}]{BARLOW1990496}%
  \BibitemOpen
  \bibfield  {author} {\bibinfo {author} {\bibfnamefont {R.}~\bibnamefont
  {Barlow}},\ }\bibfield
  {journal} {\bibinfo  {journal} {\href {\doibase 10.1016/0168-9002(90)91334-8} {Nucl. Instrum. Methods Phys. Res.}, Sect. A}\
  }\textbf {\bibinfo {volume} {297}},\ \bibinfo {pages} {496} (\bibinfo {year}
  {1990})\BibitemShut {NoStop}%
\bibitem [{\citenamefont {Verkerke}\ and\ \citenamefont
  {Kirkby}(2003)}]{Verkerke:2003ir}%
  \BibitemOpen
  \bibfield  {author} {\bibinfo {author} {\bibfnamefont {W.}~\bibnamefont
  {Verkerke}}\ and\ \bibinfo {author} {\bibfnamefont {D.~P.}\ \bibnamefont
  {Kirkby}},\ }\bibfield  {booktitle} {in \emph {\bibinfo {booktitle}
  {{Statistical Problems in Particle Physics, Astrophysics and Cosmology
  (PHYSTAT 05): Proceedings}}}, Oxford, UK, September 12-15, 2005,\ }\href@noop
  {} {\bibfield  {journal} {\bibinfo  {journal} {eConf}\ }\textbf {\bibinfo
  {volume} {C 0303241}},\ \bibinfo {pages} {MOLT007} (\bibinfo {year} {2003})},\
  \Eprint {http://arxiv.org/abs/physics/0306116} {arXiv:physics/0306116}
  \BibitemShut {NoStop}%
\end{thebibliography}
\providecommand{\noopsort}[1]{}\providecommand{\singleletter}[1]{#1}%

\end{document}